\documentclass[showpacs,aps,graphicx,]{revtex4}
\usepackage{amsmath}
\usepackage{mathrsfs}
\usepackage{amsfonts}
\usepackage{amssymb}
\usepackage{graphicx}
\usepackage{caption}
\usepackage{eufrak}
\usepackage{multirow}
\usepackage{float}
\usepackage[colorlinks,linkcolor=blue,anchorcolor=blue,citecolor=blue]{hyperref}
\usepackage{soul,color,xcolor}
\soulregister\upcite7
\soulregister\citep7
\soulregister\citet7
\soulregister\ref7
\soulregister\pageref7
\soulregister\and7
\soulregister\Name7

\begin{document}

\title{Synthesis and upper bound of Schmidt rank of the bipartite controlled-unitary gates}

\author{Gui-Long Jiang\textsuperscript{1}, Hai-Rui Wei\textsuperscript{1}\footnote{Corresponding author: hrwei@ustb.edu.cn},  Guo-Zhu Song\textsuperscript{2}, and  Ming Hua\textsuperscript{3} }

\address{1 School of Mathematics and Physics, University of Science and Technology Beijing, Beijing 100083, China \\
%
%
2 College of Physics and Materials Science, Tianjin Normal University, Tianjin 300387, China\\
3 Department of Applied Physics, School of Physical Science and Technology, Tiangong University, Tianjin, 300387 China}

\date{\today }

\begin{abstract}

Quantum circuit model is the most popular paradigm for implementing complex quantum computation.
Based on Cartan decomposition, we show that $2(N-1)$ generalized controlled-$X$ (GCX) gates, $6$ single-qubit rotations about the $y$- and $z$-axes, and $N+5$ single-partite $y$- and $z$-rotation-types which are defined in this paper are sufficient to simulate a controlled-unitary gate $\mathcal{U}_{cu(2\otimes N)}$ with $A$ controlling on $\mathbb{C}^2\otimes \mathbb{C}^N$.
In the scenario of the unitary gate $\mathcal{U}_{cd(M\otimes N)}$ with $M\geq3$ that is locally equivalent to a diagonal unitary on $\mathbb{C}^M\otimes \mathbb{C}^N$, $2M(N-1)$ GCX gates and $2M(N-1)+10$ single-partite $y$- and $z$-rotation-types are required to simulate it.
The quantum circuit for implementing $\mathcal{U}_{cu(2\otimes N)}$ and $\mathcal{U}_{cd(M\otimes N)}$ are presented.
Furthermore, we find $\mathcal{U}_{cu(2\otimes2)}$ with $A$ controlling has Schmidt rank two, and in other cases the diagonalized form of the target unitaries can be expanded in terms of specific simple types of product unitary operators.


Keywords: quantum circuit, controlled-unitary gate, quantum Schmidt rank,

\end{abstract}

\pacs{03.67.Lx, 03.65.Ud, 03.67.Mn}

\maketitle

\newcommand{\upcite}[1]{\textsuperscript{\textsuperscript{\cite{#1}}}}

\section{Introduction}\label{sec1}

Tremendous progress has been made in quantum computation\upcite{comput1,comput2,comput3,comput4} and communication\upcite{commun1,commun2,commun3,commun4,commun5} in recent years.
Unitary operations play a central role in many quantum information processing tasks: quantum circuits,\upcite{universal1,universal2,universal3,universal4,universal5,universal6,universal7} quantum algorithm,\upcite{algorithm1} creating quantum entangled states,\upcite{cluster1,cluster2,cluster3,state4} quantum state fusion,\upcite{fusion1,fusion2} cryptography,\upcite{cryptography} entanglement purification and concentration,\upcite{purifi1,purifi2,purifi3}  etc. There are local and nonlocal unitary operations. The local unitary operations, known as the tensor product operators locally acting on subsystems, can be deterministically implemented by local operations and classical communication (LOCC).\upcite{Soeda} Unfortunately, the local unitary operations alone cannot create entanglement.  The nonlocal unitary operations, which can not be implemented by LOCC, have a more complex structure and  play a more powerful role than local unitary operations in quantum information processing.\upcite{Cao1,Cao2,Cao3} So far, the properties and implementations of nonlocal unitary operations, even the simplest and the most popular controlled-unitary operations, are still far from complete. Hence it is interesting and important to find simple ways to implement nonlocal unitary operations.

There are two ways to study complex unitary operations. One approach is to decompose a unitary operation into a sum of product operations:
$\mathcal{U}=\sum_{j=1}^{m} A_j \otimes  B_j \otimes  C_j \otimes \cdots\otimes R_j$, where $A_j$, $B_j$, etc. are linearly independent local operations acting on respective parties.
The smallest possible $m$ is defined as Schmidt rank $\text{Sch}(\mathcal{U})$,\upcite{rank} which can be used for quantifying the nonlocality of $\mathcal{U}$ defined by $K_{\text{Har}}(\mathcal{U})\equiv \log_2[\text{Sch}(\mathcal{U})]$,\upcite{Power2003,Yu2016-1,Yu2016-2} providing a sufficient condition for when $\mathcal{U}$ is a controlled-unitary operation,\upcite{Yu2013,Yu2014-1,Yu2014-2,2-qubit-Rank1,Yu2015,Yu2021,Yu2010} and optimizing the synthesis of quantum computation\upcite{gate} and quantum transistors\upcite{Cao2}.
Another common approach, called synthesis or quantum circuit, is to factorize $\mathcal{U}$ into fewer achievable simple local and nonlocal operations from a universal library, which may simplify such physical implementations: $\mathcal{U}= X_1 X_2 \cdots  X_m$.  The complexity of the quantum circuit is characterized by assessing the number of entangled operations involved in the implementation. Quantum circuit is the dominant paradigm for implementing efficiently complex quantum computation.

One of the central problems in quantum computing is to minimize the number of one- and two-partite gates required to implement a desired quantum gate.
Utilizing Cartan decomposition,\upcite{Cartan} Vatan et al.\upcite{optimal} designed a controlled-NOT (CNOT)-optimized general two-qubit quantum circuit in 2004, Shende et al.\upcite{lower} presented the highest known lower bound on asymptotic CNOT cost required to implement an unstructured $n$-qubit quantum computing, Di and Wei\upcite{Di1,Di2} synthesized universal multiple-valued quantum  circuits. Using higher-dimensional Hilbert spaces, Lanyon et al.\upcite{comput1} reduced the cost of a Toffoli gate from six CNOTs to three CNOTs in 2009,
Li et al.\upcite{Liwendong} further optimize the $n$-qubit universal quantum circuit, Liu and Wei\upcite{universal2} decreased the complexity of a Fredkin gate to three entangling gates in 2020. Nowadays, many works have been devoted to multiple-valued quantum circuit\upcite{Di1,Di2}. Nonetheless, the synthesis of multi-valued quantum gates is still far from complete, and multiple-valued unitary operations still are open problems.

Bipartite controlled-unitary operation is one of the most easily understood,  extensively studied, and widely used quantum operations, such as CNOT gate.
In this paper, we study the synthesis and possible Schmidt rank of the controlled-unitary operation on bipartite Hilbert space.
By analogy with single-qubit rotations, we give the concepts of single-partite $y$- and $z$-rotation-types in higher-dimensional system, and present the synthesis of arbitrary single-partite unitary gate.
Then, utilizing Cartan decomposition technique, we present a program for synthesizing a bipartite controlled-unitary gate with $A$ controlling on $\mathbb{C}^2\otimes \mathbb{C}^N$ with $2(N-1)$ generalized controlled-$X$ gates (GCX), $6$ single-qubit rotations about the $y$- and $z$-axes, and $N+5$ single-partite $y$- and $z$-rotation-types in the worst case.
And generally, the gate on $\mathbb{C}^M\otimes \mathbb{C}^N(M\geq3)$ locally equivalent to a diagonal gate, which is the subset of controlled-unitaries with $A$ or $B$ controlling, was synthesized by $2M(N-1)$ GCX gates and $2M(N-1)+10$ single-partite $y$- and $z$-rotation-types in the worst case. Furthermore, the possible Schmidt rank of the bipartite controlled-unitary operation are presented in detail. The results indicate that upper bound of Schmidt rank of the bipartite controlled-unitary operation with $A$ controlling on $\mathbb{C}^2\otimes \mathbb{C}^N$ is $2N$, and the one on $\mathbb{C}^2\otimes \mathbb{C}^2$ is two. In the scenario of the unitary operation on $\mathbb{C}^M\otimes \mathbb{C}^N$ locally equivalent to a diagonal unitary, the upper bound is $MN$.

\section{Cartan decomposition of Lie group U(N)} \label{sec2}

Cartan decomposition technique is a extremely valuable tool employed to obtain the possible Schmidt rank and design compact quantum circuits.\upcite{Power2003,optimal,synthesis2,epl}
Cartan decomposition of real semi-simple Lie algebra $\mathfrak{g}$ is defined as
\begin{eqnarray}        \label{eq1}
\mathfrak{g}=\mathfrak{l} \oplus \mathfrak{p}.
\end{eqnarray}
The Lie subalgebra $\mathfrak{l}$ and the complement subspace $\mathfrak{p} = \mathfrak{l}^\perp$ satisfying the commutation relations
\begin{eqnarray}        \label{eq2}
[\mathfrak{l},\mathfrak{l}]\subseteq \mathfrak{l},\;\;
[\mathfrak{l},\mathfrak{p}]\subseteq \mathfrak{p},\;\;
[\mathfrak{p},\mathfrak{p}]\subseteq \mathfrak{l}.
\end{eqnarray}
Let $G$ be a compact Lie group with a real semi-simple Lie algebra $\mathfrak{g}$, then
\begin{eqnarray}        \label{eq3}
G=K_1\cdot A \cdot K_2,
\end{eqnarray}
with $K_1$, $K_2\in e^\mathfrak{l}$ and $A\in e^\mathfrak{h}$ for the Lie group exponential.   $\mathfrak{h}$ is the maximal Abelian subalgebra of ($\mathfrak{g}$, $\mathfrak{l}$) contained in $\mathfrak{p}$.

It is well known that the generators of  the Lie algebra $u(2)$ can be given by
\begin{eqnarray}        \label{eq4}
u(2)=\textrm{span}\{\sigma_x,\sigma_y,\sigma_z, I_2\}.
\end{eqnarray}
 Here
\begin{eqnarray}        \label{eq5}
\begin{split}
\sigma_x=T_{12}^{(2,1)}=\left(
  \begin{array}{cccc}
    0 & 1  \\
    1 & 0 \\
      \end{array}
\right),\;\;
\sigma_y=T_{12}^{(2,2)}=\left(
  \begin{array}{cccc}
    0 & -\mathrm{i} \\
    \mathrm{i} & 0 \\
      \end{array}
\right),\;\;
\sigma_z=T_{12}^{(2,3)}=\left(
  \begin{array}{cccc}
    1 & 0  \\
    0 & -1 \\
      \end{array}
\right),\;\;
I_2=\left(
  \begin{array}{cccc}
    1 & 0  \\
    0 & 1 \\
      \end{array}
\right).
\end{split}
\end{eqnarray}
The basis for $u(3)$ takes the form
\begin{eqnarray}        \label{eq6}
\begin{split}
&
T_{12}^{(3,1)}=\left(
  \begin{array}{cccc}
    0 & 1 & 0\\
    1 & 0 & 0\\
    0 & 0 & 0\\
  \end{array}
\right),\;\;\;\;
T_{13}^{(3,1)}=\left(
  \begin{array}{cccc}
    0 & 0 & 1\\
    0 & 0 & 0\\
    1 & 0 & 0\\
  \end{array}
\right),\;\;\;
T_{23}^{(3,1)}=\left(
  \begin{array}{cccc}
    0 & 0 & 0\\
    0 & 0 & 1\\
    0 & 1 & 0\\
  \end{array}
\right),\\
&
T_{12}^{(3,2)}=\left(
  \begin{array}{cccc}
    0            & -\mathrm{i} & 0\\
    \mathrm{i}   & 0           & 0\\
    0            & 0            & 0\\
  \end{array}
\right),\;\;
T_{13}^{(3,2)}=\left(
  \begin{array}{cccc}
    0 & 0 & -\mathrm{i}\\
    0 & 0 & 0\\
    \mathrm{i} & 0 & 0\\
  \end{array}
\right),\;
T_{23}^{(3,2)}=\left(
  \begin{array}{cccc}
    0 & 0          & 0\\
    0 & 0          & -\mathrm{i}\\
    0 & \mathrm{i} & 0\\
  \end{array}
\right),\\
&
T_{12}^{(3,3)}=\left(
  \begin{array}{cccc}
    1 & 0 & 0\\
    0 & -1 & 0\\
    0 & 0 & 0\\
  \end{array}
\right),\;
T_{13}^{(3,3)}=\left(
  \begin{array}{cccc}
    1 & 0 & 0\\
    0 & 0 & 0\\
    0 & 0 & -1\\
  \end{array}
\right),\;
I_3=\left(
  \begin{array}{cccc}
    1 & 0 & 0\\
    0 & 1 & 0\\
    0 & 0 & 1\\
  \end{array}
\right).
\end{split}
\end{eqnarray}
Lie algebra $u(N)$ is spanned by $N^2$ matrices $\{T_{ab}^{(N,1)}$, $T_{ab}^{(N,2)}$, $T_{1a}^{(N,3)}, I_N\}$ $ (a<b)$, where $I_N$ is the $N\times N$ identity matrix, $T_{ab}^{(N,1)}$, $T_{ab}^{(N,2)}$, and $T_{1a}^{(N,3)}$ are defined as
\begin{eqnarray}        \label{eq7}
(T_{ab}^{(N,1)})_{cd}=
\left\{
             \begin{array}{ccc}
             1,       &  \;c=a, d=b  &     \\
             1,       &  \;c=b, d=a  &   \\
             0,       &  \text{others} &  \\
             \end{array}
\right.
\end{eqnarray}
\begin{eqnarray}        \label{8}
(T_{ab}^{(N,2)})_{cd}=
\left\{
             \begin{array}{ccc}
             -\mathrm{i},       &  \;c=a, d=b  &     \\
             \mathrm{i},       &  \;c=b, d=a  &   \\
             0,       &  \text{others} &  \\
             \end{array}
\right.
\end{eqnarray}
\begin{eqnarray}        \label{eq9}
(T_{1a}^{(N,3)})_{cd}=
\left\{
             \begin{array}{ccc}
             1,       &  c=d=1  &     \\
             -1,      &  c=d=a, &  2\leq a\leq N. \\
             0,       &  \text{others} &  \\
             \end{array}
\right.
\end{eqnarray}
The subscript $a$ and $b$ are order index of the basis. The $c$ and $d$ represent the $c$th row and $d$th column in matrix $T_{ab}^{(N,1)}$, $T_{ab}^{(N,2)}$, and  $T_{1a}^{(N,3)}$. The $b,c,d\in[1, 2, \cdots, N]$, and $N$ represents the $N$ dimensional system.

Based on Equations (\ref{eq1}-\ref{eq9}), it is easily to find that the Cartan decomposition of Lie algebra $u(N)$ may has the form $u(N)=\mathfrak{l}_{u(N)} \oplus \mathfrak{p}_{u(N)}$, where
\begin{eqnarray}        \label{eq10}
\mathfrak{l}_{u(N)}=\textrm{span}\{T_{ab}^{(N,2)}, I_N\},
\end{eqnarray}
\begin{eqnarray}        \label{eq11}
\mathfrak{p}_{u(N)}=\textrm{span}\{T_{ab}^{(N,1)},T_{1a}^{(N,3)}\}.
\end{eqnarray}
The Cartan subalgebra $\mathfrak{h}_{u(N)}$ is generated by
\begin{eqnarray}        \label{eq12}
\mathfrak{h}_{u(N)}=\textrm{span}\{T_{1a}^{(N,3)}\}.
\end{eqnarray}
Then, the associated Cartan decomposition of group $\mathcal{U}_{u(N)}\in U(N)$ is of the form
\begin{eqnarray}        \label{eq13}
\begin{split}
\mathcal{U}_{u(N)}=& e^{\mathrm{i}\theta_N}\cdot M_1 \cdot e^{\mathrm{i}(\theta_1 T_{12}^{(N,3)}+ \theta_2 T_{13}^{(N,3)}+,\cdots,+\theta_{N-1} T_{1N}^{(N,3)})}\cdot M_2,
\end{split}
\end{eqnarray}
with $M_1,\;M_2\in e^{\textrm{span}\{T_{ab}^{(N,2)}\}}$.
For instance, any two-dimensional unitary operator can be expanded into
\begin{eqnarray}        \label{eq}
\mathcal{U}_{u(2)}=& e^{\mathrm{i}\theta_2} \cdot
e^{\mathrm{i} \vartheta_1 \sigma_y} \cdot
e^{\mathrm{i}\theta_1 \sigma_z} \cdot
e^{\mathrm{i}\widetilde{\vartheta}_1 \sigma_y}.
\end{eqnarray}
Any three-dimensional unitary operator can be expanded into
\begin{eqnarray}        \label{eq14}
\mathcal{U}_{u(3)}=& e^{\mathrm{i}\theta_3}\cdot e^{\mathrm{i}(\vartheta_1 T_{12}^{(3,2)}+\vartheta_2 T_{13}^{(3,2)}+ \vartheta_{3} T_{23}^{(3,2)})} \cdot
e^{\mathrm{i}(\theta_1 T_{12}^{(3,3)}+ \theta_2 T_{13}^{(3,3)})} \cdot
e^{\mathrm{i}(\widetilde{\vartheta}_1 T_{12}^{(3,2)}+\widetilde{\vartheta}_2 T_{13}^{(3,2)}+ \widetilde{\vartheta}_{3} T_{23}^{(3,2)})}.
\end{eqnarray}

Local single-qubit gates $e^{\mathrm{i}\vartheta\sigma_y}$ ($e^{\mathrm{i}\widetilde{\vartheta}\sigma_y}$) and $e^{\mathrm{i}\beta\sigma_z}$ are single-qubit rotations about the $y$- and $z$-axes, respectively.
That is, two single-qubit rotations about the $y$-axes and one single-qubit rotation about the $z$-axes are sufficient to simulate a generic single-qubit gate.
The global phase factor $ e^{\mathrm{i}\theta_2}$ is missing here because it is irrelevant for quantum information processing.
The matrices in the set $\{T_{ab}^{^{(N,2)}}\}$ $(a<b)$ do not commute with each other.

\textbf{Definition}:
Similar to the definition of single-qubit rotations, we define
$e^{\mathrm{i}(\vartheta_1 T_{12}^{(N,2)}+\vartheta_2 T_{13}^{(N,2)}+\cdots+ \vartheta_{N(N-1)/2} T_{(N-1)N}^{(N,2)})}$
($e^{\mathrm{i}(\widetilde{\vartheta}_1 T_{12}^{(N,2)}+\widetilde{\vartheta}_2 T_{13}^{(N,2)}+\cdots+ \widetilde{\vartheta}_{N(N-1)/2} T_{(N-1)N}^{(N,2)})}$) as single-partite $y$-rotation-type, and
$e^{\mathrm{i}(\theta_1 T_{12}^{(N,3)}+ \theta_2 T_{13}^{(N,3)}+,\cdots,+\theta_{N-1} T_{1N}^{(N,3)})}$
as single-partite $z$-rotation-type. When $N=2$, they are compatible with single-qubit rotations about the $y$- and $z$-axes.

Based on Equations (\ref{eq10}-\ref{eq13}), one can see that arbitrary single-partite multi-valued unitary $\mathcal{U}_{u(N)}$ can be implemented by two single-partite $y$-rotation-types and one $z$-rotation-type.  Each $y$-rotation-type and $z$-rotation-type has $N(N-1)/2$ and $N-1$ free parameters, respectively. We omit the overall phase change $ e^{\mathrm{i}\theta_N}$.

\section{Synthesis and quantum Schmidt rank of controlled-unitary gate with $A$ controlling on $\mathbb{C}^2\otimes \mathbb{C}^N$}\label{sec3}

It is known that, if the controlled-unitary gate $\mathcal{U}_{cu(M\otimes N)}$ on $\mathbb{C}^2\otimes \mathbb{C}^N$ is controlled from $A$ side, then $\exists U_A$, $V_A$ unitaries such that
\begin{eqnarray}        \label{eq16}
\begin{split}
\mathcal{U}_{cu(M\otimes N)}=(U_A\otimes I_B) \cdot (\sum_{i=0}^{M-1} |i\rangle_A\langle i| \otimes U_i) \cdot (V_A\otimes I_B).
\end{split}
\end{eqnarray}
That is, up to local unitaries, $\mathcal{U}_{cu(M\otimes N)}$  with $A$ controlling is equivalent to $\widetilde{\mathcal{U}}_{cu(M\otimes N)}=\sum_{i=0}^{M-1} |i\rangle_A\langle i| \otimes U_i$.

Using the well-known commutation relations
\begin{eqnarray}        \label{eq17}
[A \otimes B, C \otimes D]= [A,C]\otimes (B\cdot D)+ (C \cdot A)\otimes [B,D],
\end{eqnarray}
one can verify that the Cartan decomposition of $\widetilde{\mathcal{U}}_{cu(2\otimes N)}$ has the form $\widetilde{u}_{cu(2\otimes N)}=\mathfrak{l}_{u(2\otimes N)}\oplus\mathfrak{p}_{u(2\otimes N)}$,
where
\begin{eqnarray}        \label{eq18}
\mathfrak{l}_{u(2\otimes N)}=\textrm{span}\{I_2 \otimes u(N), \sigma_z\otimes I_N\},
\end{eqnarray}
\begin{eqnarray}        \label{eq19}
\mathfrak{p}_{u(2\otimes N)}=\textrm{span}\{\sigma_z\otimes su(N)\}
\end{eqnarray}
with $su(N)=\textrm{span}\{T_{ab}^{(N,1)}, T_{ab}^{(N,2)}, T_{1a}^{(N,3)}\}$.
And the Cartan subalgebra is
\begin{eqnarray}        \label{eq20}
\mathfrak{h}_{u(2\otimes N)}=\textrm{span}\{\sigma_z\otimes T_{1a}^{(N,3)}\}.
\end{eqnarray}
As a consequence, $\mathcal{U}_{cu(2\otimes N)}$ with $A$ controlling can be decomposed as
\begin{eqnarray}        \label{eq21}
\mathcal{U}_{cu(2\otimes N)}=(U_A\otimes U_B) \cdot\wedge(\triangle_{2\otimes N})\cdot (V_A\otimes V_B),
\end{eqnarray}
where $\wedge(\triangle_{2\otimes N})=e^{\textrm{span}\{\sigma_z\otimes T_{1a}^{(N,3)}\}}.$

Based on Equation (\ref{eq13}), it is easily verified that each of local unitary operators $U_A$ and $V_A$ acting on subsystem $A$ can be implemented by three single-partite rotations about the $y$- and $z$-axes.
Each of local unitary operators $U_B$ and $V_B$ acting on subsystem $B$ can  be implemented by three single-partite $y$- and $z$-rotation-types.

\begin{figure}[!h]
\begin{center}
 \includegraphics[width=7.0 cm]{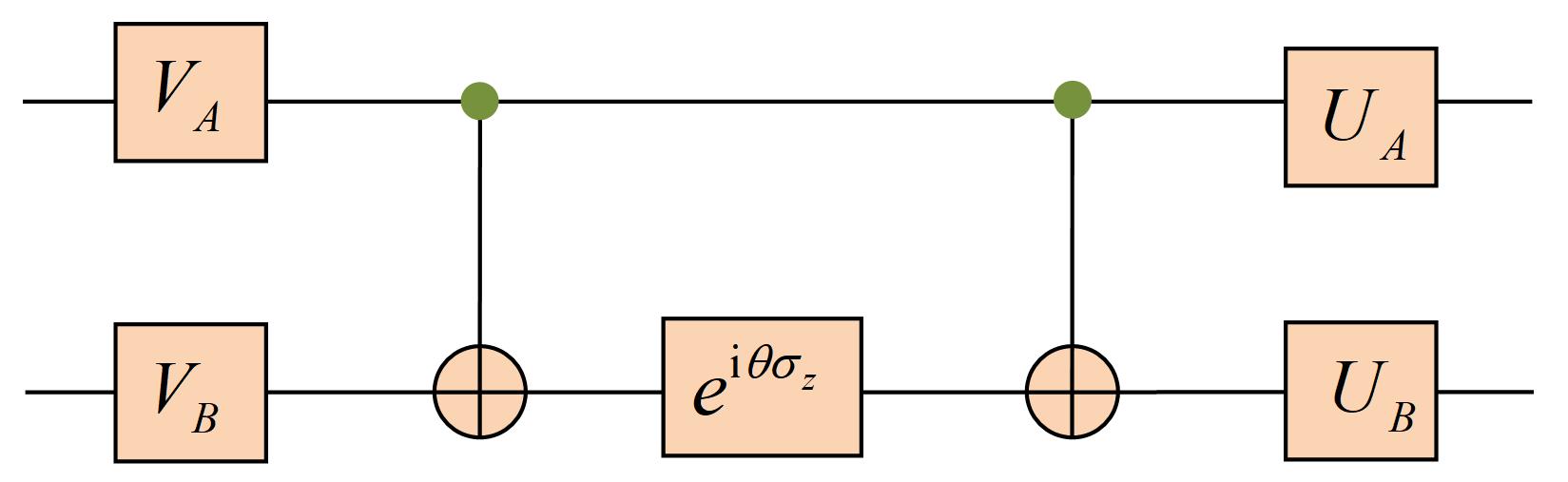}
  \caption{Synthesis of a controlled-unitary gate with $A$ controlling on $\mathbb{C}^2\otimes \mathbb{C}^2$. Each of local operations $U_A$, $V_A$, $U_B$, and $V_B$ is the product of three single-qubit rotations about the $y$- and $z$-axes. The overall phase change is ignored here because it has no effect on the present program.}
  \label{Fig22}
\end{center}
\end{figure}

\emph{Synthesis and Schmidt rank of $\mathcal{U}_{cu(2\otimes2)}$ with $A$ controlling}.
When $\mathcal{U}_{cu}$ with $A$ controlling is acting on $\mathbb{C}^2\otimes \mathbb{C}^2$, it can be decomposed as
\begin{eqnarray}        \label{eq22}
\begin{split}
\mathcal{U}_{cu(2\otimes2)}&=(U_A\otimes U_B)e^{\mathrm{i}\theta\sigma_z \otimes \sigma_z} (V_A\otimes V_B)\\
                           &=(U_A\otimes U_B)\cdot \text{CNOT} \cdot e^{\mathrm{i}\theta I_2 \otimes \sigma_z} \cdot \text{CNOT} \cdot(V_A\otimes V_B).
\end{split}
\end{eqnarray}
Here each of single-qubit gates $U_A$, $V_A$, $U_B$, and $V_B$ can be expanded in the form of $e^{\mathrm{i}\alpha\sigma_y} \cdot e^{\mathrm{i}\beta\sigma_z} \cdot e^{\mathrm{i}\widetilde{\alpha}\sigma_y}$, that is, each one is the product of three single-qubit rotations about the $y$- and $z$-axes, see Equation (\ref{eq}). CNOT represents a controlled-NOT gate, and it is given by the matrix
\begin{eqnarray}        \label{eq23}
\text{CNOT}=\left(
  \begin{array}{cccc}
    1 & 0 & 0 & 0 \\
    0 & 1 & 0 & 0 \\
    0 & 0 & 0 & 1 \\
    0 & 0 & 1 & 0 \\
  \end{array}
\right).
\end{eqnarray}
Therefore, as shown in Figure \ref{Fig22}, two CNOT gates together with $13=3\times4+1$ single-qubit rotations about the $y$- and $z$-axes are sufficient to simulate a controlled-unitary $\mathcal{U}_{cu(2\otimes2)}$ with $A$ controlling.

Based on Equation (\ref{eq22}), we may expand $\wedge(\triangle_{2\otimes 2})=e^{\mathrm{i}\theta\sigma_z\otimes \sigma_z}$ as
\begin{eqnarray}        \label{eq24}
\begin{split}
\wedge(\triangle_{2\otimes 2})=c_{\theta}I_2\otimes I_2 + \mathrm{i}s_{\theta} \sigma_z\otimes \sigma_z,
\end{split}
\end{eqnarray}
where $c_{\theta}\equiv\cos(\theta)$, $s_{\theta}\equiv\sin(\theta)$. Therefore, $\mathcal{U}_{cu(2\otimes2)}$ with $A$ controlling only has Schmidt rank 2 when $c_{\theta}$, $s_{\theta}\neq0$.

\begin{figure}[!h]
\begin{center}
  \includegraphics[width=10 cm]{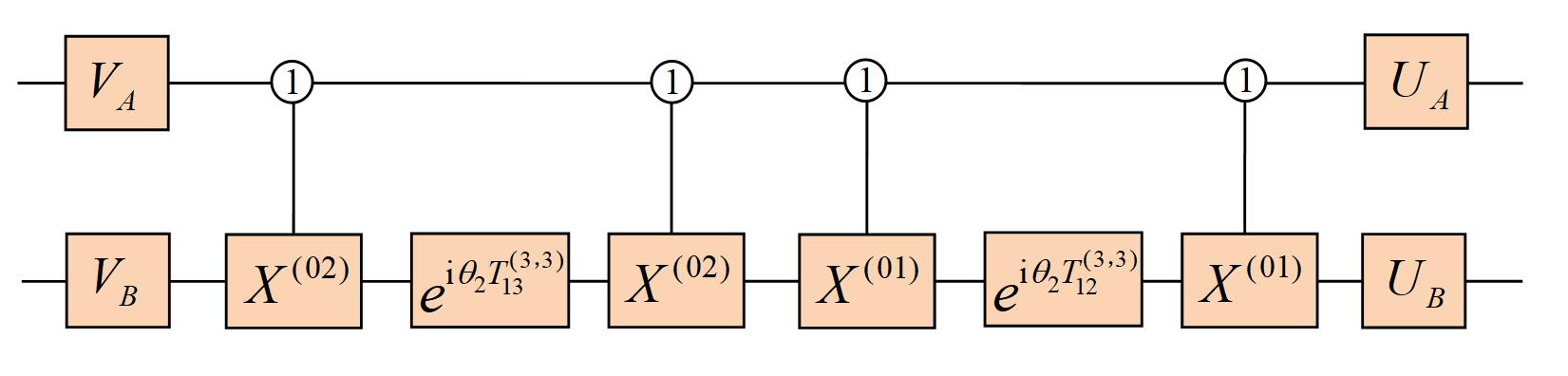}
  \caption{Synthesis of a controlled-unitary gate with $A$ controlling on $\mathbb{C}^2\otimes \mathbb{C}^3$. Both local operations $U_B$ and $V_B$ are the product of three single-partite $y$- and $z$-rotation-types from Equation (\ref{eq14}). We omit  the overall phase change.}
  \label{Fig23-NEW}
\end{center}
\end{figure}

\begin{figure}[!h]
\begin{center}
  \includegraphics[width=13.5 cm]{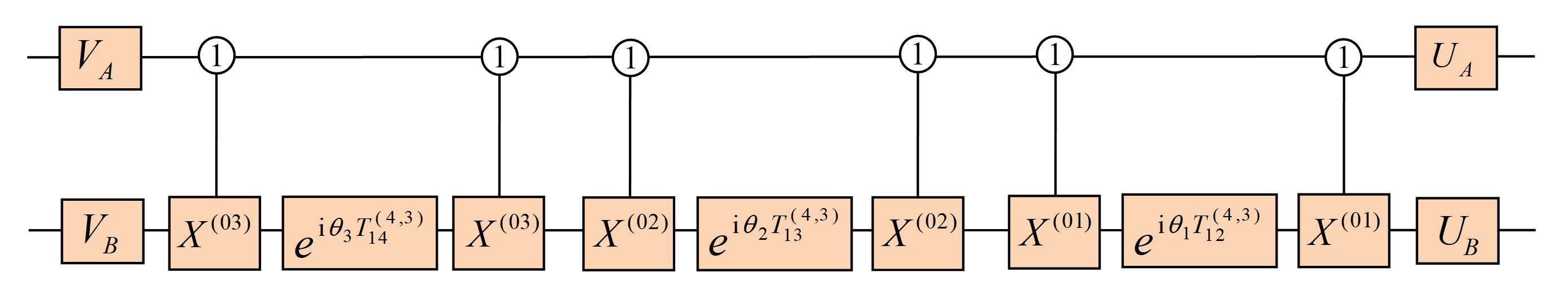}
  \caption{Synthesis of a controlled-unitary gate with $A$ controlling on $\mathbb{C}^2\otimes \mathbb{C}^4$. The global phase factor is missing here.}
  \label{Fig24-NEW}
\end{center}
\end{figure}

\emph{Synthesis and Schmidt rank of $\mathcal{U}_{cu(2\otimes3)}$ with $A$ controlling}.
When $\mathcal{U}_{cu}$ with $A$ controlling is acting on $\mathbb{C}^2\otimes \mathbb{C}^3$, it can be decomposed as
\begin{eqnarray}        \label{eq25}
\begin{split}
\mathcal{U}_{cu(2\otimes3)}=&(U_A\otimes U_B)e^{\mathrm{i} (\theta_1\sigma_z\otimes T_{12}^{(3,3)}+\theta_2 \sigma_z\otimes T_{13}^{(3,3)})} (V_A\otimes V_B).
\end{split}
\end{eqnarray}
We find that each $e^{\mathrm{i}(\theta\sigma_z\otimes T_{1a}^{(3,3)})}$  can be synthesized by
\begin{eqnarray}        \label{eq26}
\begin{split}
e^{\mathrm{i} (\theta\sigma_z\otimes T_{1a}^{(3,3)})}
                = \text{GCX}(1\rightarrow X^{(0,a-1)}) \cdot e^{\mathrm{i}\theta I_2 \otimes T_{1a}^{(3,3)}} \cdot \text{GCX}(1\rightarrow X^{(0,a-1)})).
\end{split}
\end{eqnarray}
Here the universal generalized controlled-$X$ gate,\textsuperscript{\cite{Di1,Di2}} GCX$(m\rightarrow X^{(ij)})$, implements the operation
\begin{eqnarray}        \label{eq27}
X^{(ij)}=|i\rangle\langle j|+ |j\rangle\langle i|+\sum_{k\neq i,j} |k\rangle\langle k|,
\end{eqnarray}
on the target particle if and only if the control particle is in the state $|m\rangle$,  and has no effect otherwise. Hence, Equation (\ref{eq25}) can be rewritten as
\begin{eqnarray}        \label{eq28}
\begin{split}
\mathcal{U}_{cu(2\otimes3)}=&(U_A\otimes U_B)\cdot
                 \text{GCX}(1\rightarrow X^{(01)}) \cdot e^{\mathrm{i}\theta_1 I_2 \otimes T_{12}^{(3,3)}} \cdot \text{GCX}(1\rightarrow X^{(01)})\\ &\cdot
                 \text{GCX}(1\rightarrow X^{(02)}) \cdot e^{\mathrm{i}\theta_2 I_2 \otimes T_{13}^{(3,3)}} \cdot \text{GCX}(1\rightarrow X^{(02)})\cdot
                   (V_A\otimes V_B).
\end{split}
\end{eqnarray}
That is to say, a $\mathcal{U}_{cu(2\otimes3)}$ with $A$ controlling can be synthesized by four GCX gates, $6=3\times2$ single-qubit rotations about the $y$- and $z$-axes,  and $8=3\times2+2$ single-partite $y$- and $z$-rotation-types in the worst case, see Figure \ref{Fig23-NEW}.
Note that  each above $y$-rotation-type and $z$-rotation-type has 3 and 2 free parameters, respectively, see Equation (\ref{eq14}).
By the same argument as that made for $\mathbb{C}^2\otimes \mathbb{C}^4$, we can find that as shown in Figure \ref{Fig24-NEW}, the quantum circuit for implementing $\mathcal{U}_{cu(2\otimes4)}$ with $A$ controlling contains six GCX gates, $6=3\times2$ single-qubit rotations about the $y$- and $z$-axes, and $9=3\times2+3$ single-partite $y$- and $z$-rotation-types.

Based on Equation (\ref{eq25}), we may expand $\wedge(\triangle_{2\otimes 3})=e^{\mathrm{i}(\theta_1\sigma_z\otimes T_{12}^{(3,3)}+ \theta_2\sigma_z\otimes T_{13}^{(3,3)})}$ as
\begin{eqnarray}        \label{eq29}
\begin{split}
\wedge(\triangle_{2\otimes 3})
&=\frac{1}{3}(c_{(\theta_1+\theta_2)}+c_{\theta_1}+c_{\theta_2})I_{6}+\frac{1}{3}(c_{(\theta_1+\theta_2)}-2c_{\theta_1}+c_{\theta_2})I_{2}\otimes T_{12}^{(3,3)}\\
&+\frac{1}{3}(c_{(\theta_1+\theta_2)}+c_{\theta_1}-2c_{\theta_2})I_{2}\otimes T_{13}^{(3,3)}+\frac{1}{3}\mathrm{i}(s_{(\theta_1+\theta_2)}-s_{\theta_1}-s_{\theta_2})\sigma_z\otimes I_{3}\\
&+\frac{1}{3}\mathrm{i}(s_{(\theta_1+\theta_2)}+2s_{\theta_1}-s_{\theta_2})\sigma_z\otimes T_{12}^{(3,3)}+\frac{1}{3}\mathrm{i}(s_{(\theta_1+\theta_2)}-s_{\theta_1}+2s_{\theta_2})\sigma_z\otimes T_{13}^{(3,3)},
\end{split}
\end{eqnarray}
where $s_{(\theta_1+\theta_2)} \equiv \sin(\theta_1+\theta_2)$, $c_{(\theta_1+\theta_2)} \equiv \cos(\theta_1+\theta_2)$.
That is to say, the upper bound of Schmidt rank of $\mathcal{U}_{cu(2\otimes 3)}$ with $A$ controlling is six.
By calculating similarly to Equation (\ref{eq29}), we can further find $\mathcal{U}_{cu(2\otimes 4)}$ with $A$ controlling has Schmidt rank at most eight.
The upper bound on the Schmidt rank is not tight. A tight upper bound for the Schmidt rank is only two for generic diagonal unitaries on $2\times N$ space.  But the expansion given above is interesting in that it is in terms of product unitary operators, where the local operators are either the identity operator or with support in two-dimensional subspaces.

\begin{figure}[!h]
\begin{center}
  \includegraphics[width=12.5 cm]{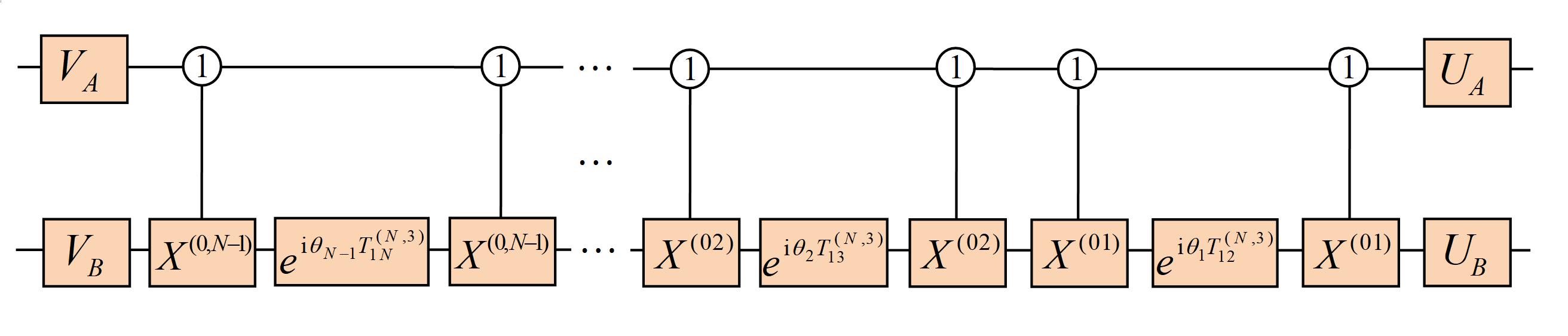}
  \caption{Synthesis of a controlled-unitary gate with $A$ controlling on $\mathbb{C}^2\otimes \mathbb{C}^N$. The global phase  is omitted.}
  \label{Fig2N}
\end{center}
\end{figure}

\emph{Synthesis and Schmidt rank of $\mathcal{U}_{cu(2\otimes N)}$ with $A$ controlling}.
When $\mathcal{U}_{cu}$ is acting on $\mathbb{C}^2\otimes \mathbb{C}^N$, it can be decomposed as
\begin{equation}        \label{eq30}
\begin{split}
\mathcal{U}_{cu(2\otimes N)}=&(U_A\otimes U_B) \cdot e^{\mathrm{i} (\theta_1\sigma_z\otimes T_{12}^{(N,3)}+\theta_2 \sigma_z\otimes T_{13}^{(N,3)}+\cdots+\theta_{N-1} \sigma_z\otimes T_{1N}^{(N,3)})} \cdot (V_A\otimes V_B)\\
                =&(U_A\otimes U_B)\cdot
                 \text{GCX}(1\rightarrow X^{(01)}) \cdot e^{\mathrm{i}\theta_1 I_2 \otimes T_{12}^{(N,3)}} \cdot \text{GCX}(1\rightarrow X^{(01)})\\ &\cdot
                 \text{GCX}(1\rightarrow X^{(02)}) \cdot e^{\mathrm{i}\theta_2 I_2 \otimes T_{13}^{(N,3)}} \cdot \text{GCX}(1\rightarrow X^{(02)})\\&
                 \cdot \cdots \\&\cdot
                 \text{GCX}(1\rightarrow X^{(0,N-1)}) \cdot e^{\mathrm{i}\theta_{N-1} I_2 \otimes T_{1N}^{(N,3)}} \cdot \text{GCX}(1\rightarrow X^{(0,N-1)})\cdot                   (V_A\otimes V_B).
\end{split}
\end{equation}
We note that there are $N-1$ items in the exponential $e^{\mathrm{i} (\theta_1\sigma_z\otimes T_{12}^{(N,3)}+\cdots+\theta_{N-1} \sigma_z\otimes T_{1N}^{(N,3)})}$, and each of them can be synthesized by two GCX gates and one single-partite $z$-rotation-type.
Hence, the quantum circuit for implementing $\mathcal{U}_{cu(2\otimes N)}$ with $A$ controlling contains $2(N-1)$ GCX gates, $6=3\times2$ single-qubit rotations about the $y$- and $z$-axes, and $N+5=3\times2+(N-1)$  single-partite $y$- and $z$-rotation-types, see Figure \ref{Fig2N}. It is noted that each $y$-rotation-type and $z$-rotation-type has $N(N-1)/2$ and $N-1$ free parameters, respectively.

Based on Equation (\ref{eq30}), we may expand $\wedge(\triangle_{2\otimes N})=e^{\mathrm{i} (\theta_1\sigma_z\otimes T_{12}^{(N,3)}+\theta_2 \sigma_z\otimes T_{13}^{(N,3)}+\cdots+\theta_{N-1} \sigma_z\otimes T_{1N}^{(N,3)})}$ as
\begin{equation}        \label{eq31}
\begin{split}
\wedge(\triangle_{2\otimes N})=&\gamma_{1} I_{2N}+ \delta_{1} \sigma_{z} \otimes I_{N}
+\sum_{k=2}^{N-1}  (\gamma_{k} I_{2} \otimes T_{1k}^{(N,3)} + \delta_{k} \sigma_{z} \otimes T_{1k}^{(N,3)}).
\end{split}
\end{equation}
Here parameters $\gamma_k$ and $\delta_k$ can be expressed by following:
\begin{equation}        \label{eq32}
\gamma_{k}=
\left\{
             \begin{array}{cc}
             \displaystyle\frac{1}{N}[c_{(\theta_{1}+\theta_{2}+\cdots+\theta_{N-1})}+\displaystyle\sum\limits_{n=1}^{N-1}c_{\theta_{n}}]       &k=1,       \\
             \displaystyle\frac{1}{N}[c_{(\theta_{1}+\theta_{2}+\cdots+\theta_{N-1})}+\displaystyle\sum\limits_{n=1}^{N-1}c_{\theta_{n}}-Nc_{\theta_{k-1}}]       &k=2, \cdots, N-1.  \\
             \end{array}
\right.
\end{equation}
\begin{equation}        \label{eq33}
\delta_{k}=
\left\{
             \begin{array}{cc}
             \displaystyle\frac{\mathrm{i}}{N}[s_{(\theta_{1}+\theta_{2}+\cdots+\theta_{N-1})}-\displaystyle\sum\limits_{n=1}^{N-1}s_{\theta_{n}}]       &k=1,       \\
             \displaystyle\frac{\mathrm{i}}{N}[s_{(\theta_{1}+\theta_{2}+\cdots+\theta_{N-1})}-\displaystyle\sum\limits_{n=1}^{N-1}s_{\theta_{n}}+Ns_{\theta_{k-1}}]       &k=2, \cdots, N-1.  \\
             \end{array}
\right.
\end{equation}
Therefore, the Schmidt rank of a $\mathcal{U}_{cu(2\otimes N)}$ with $A$ controlling is no more than $2N$.

\section{Synthesis and quantum Schmidt rank of the gate locally equivalent to diagonal unitary on $\mathbb{C}^M\otimes \mathbb{C}^N$}\label{sec4}

It is well known that, if $\mathcal{U}_{cd(M\otimes N)}$ is a unitary gate locally equivalent to a diagonal unitary on $\mathbb{C}^M\otimes \mathbb{C}^N$, then it can be viewed as a controlled unitary controlled from $A$ side or $B$ side, and $\exists U_A$, $U_B$, $V_A$, $V_B$ unitaries such that
\begin{eqnarray}        \label{eq34}
\mathcal{U}_{cd(M\otimes N)}=
(U_A \otimes U_B)\cdot (\sum_{i=1}^{d_A}\sum_{j=1}^{d_B} |i\rangle_A\langle i| \otimes |j\rangle_B\langle j|) \cdot(V_A \otimes V_B).
\end{eqnarray}
Here we use the symbol $cd$ in the subscript, rather than simply using $d$, in order to distinguish them from the diagonal unitaries $(\sum_{i=1}^{d_A}\sum_{j=1}^{d_B} |i\rangle_A\langle i| \otimes |j\rangle_B\langle j|)$. For the same unitary of this type, either party may act as the control, and the choice would not affect the decomposition.

As $(\sum_{j=1}^{d_B} |i\rangle_A\langle i| \otimes |j\rangle_B\langle j|)\in e^{\text{span}\{T_{1\tilde{a}}^{(M,3)} \otimes T_{1a}^{(N,3)}, I_M \otimes T_{1a}^{(N,3)}, T_{1\tilde{a}}^{(M,3)} \otimes I_N \}}$
and the matrices in the set $\{T_{1\tilde{a}}^{(M,3)} \otimes T_{1a}^{(N,3)}, I_M \otimes T_{1a}^{(N,3)}, T_{1\tilde{a}}^{(M,3)} \otimes I_N \}$
do not commute with each other,
local single-partite $z$-rotation-type $e^{\text{span}\{I_M \otimes T_{1a}^{(N,3)}\}}$ will be ``absorbed'' by their neighboring gate $U_B$ or $V_B$,
and $e^{\text{span}\{ T_{1\tilde{a}}^{(M,3)} \otimes I_N \}}$ will be ``absorbed'' by $U_A$ or $V_A$. Then $\mathcal{U}_{cd(M\otimes N)}$ can be rewritten as
\begin{eqnarray}        \label{eq35}
\begin{split}
\mathcal{U}_{cd(M\otimes N)}
=&(U_A \otimes U_B)\cdot\wedge(\triangle_{M\otimes N})\cdot(V_A \otimes V_B),
\end{split}\end{eqnarray}
where $\wedge(\triangle_{M\otimes N})=e^{\text{span}\{T_{1\tilde{a}}^{(M,3)}\otimes T_{1a}^{(N,3)}\}}$ with $\tilde{a}\in[2, 3, \cdots, M]$ and $a\in[2, 3, \cdots, N]$.

\emph{Synthesis and Schmidt rank of $\mathcal{U}_{cd(3\otimes 3)}$ on $\mathbb{C}^3\otimes \mathbb{C}^3$}.
When the $\mathcal{U}_{cd}$ is acting on $\mathbb{C}^3\otimes \mathbb{C}^3$, it can be decomposed as
\begin{equation}        \label{eq36}
\begin{split}
\mathcal{U}_{cd(3\otimes3)}=&(U_A\otimes U_B)e^{\mathrm{i}(\theta_1 T_{12}^{(3,3)}\otimes T_{12}^{(3,3)}
                                                +\theta_2 T_{13}^{(3,3)}\otimes T_{12}^{(3,3)}
                                                +\theta_3 T_{12}^{(3,3)}\otimes T_{13}^{(3,3)}
                                                +\theta_4 T_{13}^{(3,3)}\otimes T_{13}^{(3,3)})} (V_A\otimes V_B).
\end{split}
\end{equation}

\begin{figure}[!h]
\begin{center}
  \includegraphics[width=17.0 cm]{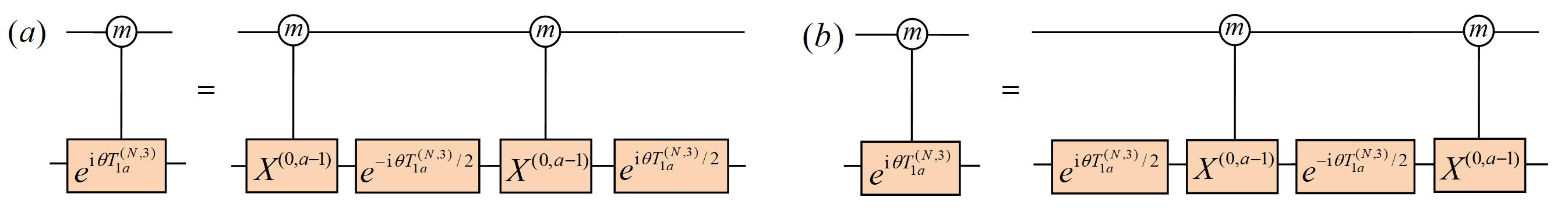}
  \caption{Synthesis of $\wedge(m\rightarrow e^{\mathrm{i} \theta T_{1a}^{(N,3)}})$ in terms of the $\text{GCX}(m\rightarrow X^{(0,a-1)})$.}
  \label{FigCM}
\end{center}
\end{figure}

\begin{figure}[!h]
\begin{center}
  \includegraphics[width=18 cm]{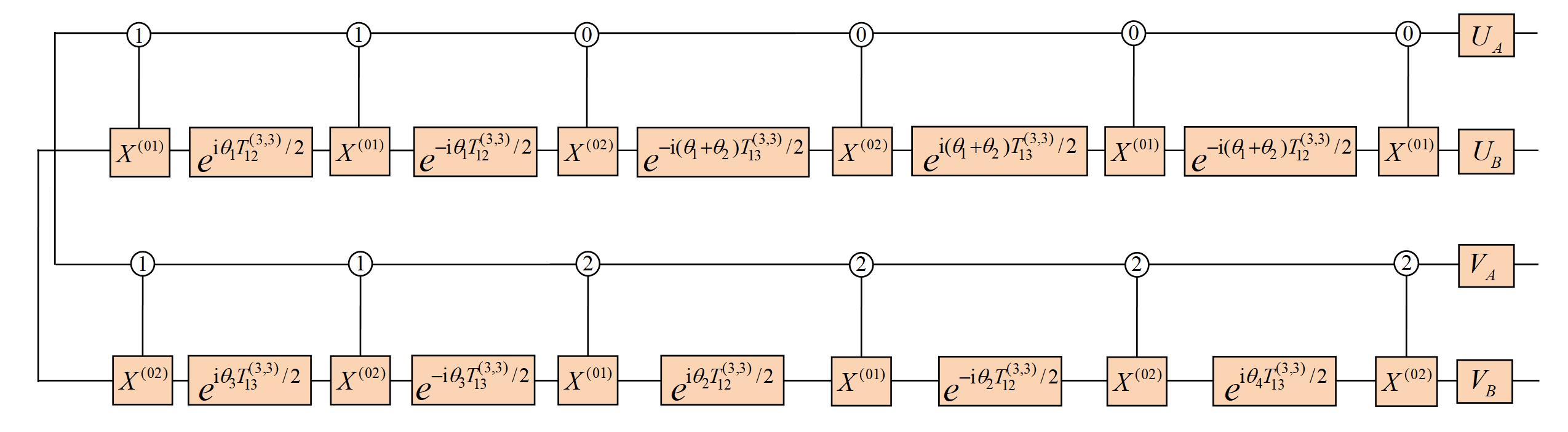}
  \caption{Synthesis for implementing a gate on $\mathbb{C}^3\otimes \mathbb{C}^3$ which is locally equivalent to a diagonal gate. Each of local operations $U_A$, $V_A$, $U_B$, and $V_B$ is the product of two single-partite $y$-rotation-types and one $z$-rotation-type from Equation (\ref{eq14}),  and the overall phase change is missing here.}
  \label{Fig33}
\end{center}
\end{figure}

One can verify that each $e^{\mathrm{i}(\theta T_{1\tilde{a}}^{(M,3)}\otimes T_{1a}^{(N,3)})}$ can be synthesized by  
\begin{equation}        \label{eq37}
\begin{split}
e^{\mathrm{i}(\theta T_{1\tilde{a}}^{(M,3)}\otimes T_{1a}^{(N,3)})}
                =&\wedge(0\rightarrow e^{\mathrm{i} \theta T_{1a}^{(N,3)}}) \cdot
                 \wedge(\tilde{a}-1\rightarrow e^{-\mathrm{i} \theta T_{1a}^{(N,3)}}),
\end{split}
\end{equation}
where $\wedge(m\rightarrow e^{\mathrm{i} \theta T_{1a}^{(N,3)}})$ is to implement the operation $e^{\mathrm{i} \theta T_{1a}^{(N,3)}}$ on the target particle if and only if the control particle is in the state $|m\rangle$, and has no effect otherwise. Given verification, $\wedge(m\rightarrow e^{\mathrm{i} \theta T_{1a}^{(N,3)}})$ can be simulated by two GCX gates and two single-partite $z$-rotation-types as described in Figure \ref{FigCM}, that is,
\begin{equation}        \label{eq38}
\begin{split}
 \wedge(m\rightarrow e^{\mathrm{i}\theta T_{1a}^{(N,3)}})=&e^{\mathrm{i} \frac{\theta}{2} I_M \otimes T_{1a}^{(N,3)}} \cdot
 \text{GCX}(m\rightarrow X^{(0,a-1)})  \cdot
 e^{-\mathrm{i} \frac{\theta}{2} I_M \otimes T_{1a}^{(N,3)}} \cdot
 \text{GCX}(m\rightarrow X^{(0,a-1)})\\
=& \text{GCX}(m\rightarrow X^{(0,a-1)})  \cdot
          e^{-\mathrm{i} \frac{\theta}{2} I_M \otimes T_{1a}^{(N,3)}} \cdot
   \text{GCX}(m\rightarrow X^{(0,a-1)})  \cdot
          e^{\mathrm{i} \frac{\theta}{2} I_M \otimes T_{1a}^{(N,3)}}.
\end{split}
\end{equation}
Hence, Equation (\ref{eq36}) can be rewritten as
\begin{equation}        \label{eq39}
\begin{split}
\mathcal{U}_{cd(3\otimes3)}
=&(U_A\otimes U_B)\cdot
                 \wedge(0\rightarrow e^{\mathrm{i}  (\theta_1+\theta_2) T_{12}^{(3,3)}}) \cdot
                 \wedge(0\rightarrow e^{\mathrm{i}  (\theta_3+\theta_4)T_{13}^{(3,3)}}) \cdot \\&
                 \wedge(1\rightarrow e^{-\mathrm{i} \theta_1T_{12}^{(3,3)}})\cdot
                 \wedge(1\rightarrow e^{-\mathrm{i} \theta_3T_{13}^{(3,3)}})\cdot
                 \wedge(2\rightarrow e^{-\mathrm{i} \theta_2T_{12}^{(3,3)}})\cdot
                 \wedge(2\rightarrow e^{-\mathrm{i} \theta_4T_{13}^{(3,3)}})\cdot
                   (V_A\otimes V_B).
\end{split}
\end{equation}
Note that the rightmost gate of the circuit of Figure \ref{FigCM}(a) will be ``absorbed'' by its neighboring gate $U_B$ of Figure \ref{Fig33}, and the leftmost gate of the circuit of Figure \ref{FigCM}(b) will be ``absorbed'' by its neighboring gate $V_B$ of Figure \ref{Fig33}.
Therefore, according to Equations (\ref{eq38}) and (\ref{eq39}), 12 GCX gates and $22=3\times4+(12-2)$ single-partite $y$- and $z$-rotation-types are sufficient to simulate a $\mathcal{U}_{cd(3\otimes3)}$ locally equivalent to a diagonal unitary as in Figure \ref{Fig33}.

Based on Equation (\ref{eq39}), we may expand $\wedge(\triangle_{3\otimes 3})$  as
\begin{equation}        \label{eq40}
\begin{split}
\wedge(\triangle_{3\otimes 3})=&CI_{3}\otimes I_{3}+[C-\frac{1}{3}(c_{(\theta_{1}+\theta_{2})}+c_{\theta_{1}}+c_{\theta_{2}})+\frac{\mathrm{i}}{3}(s_{(\theta_{1}+\theta_{2})}-s_{\theta_{1}}-s_{\theta_{2}})]I_{3}\otimes T^{(3,3)}_{12} \\
&+[C-\frac{1}{3}(c_{(\theta_{3}+\theta_{4})}+c_{\theta_{3}}+c_{\theta_{4}})+\frac{\mathrm{i}}{3}(s_{(\theta_{3}+\theta_{4})}-s_{\theta_{3}}-s_{\theta_{4}})]I_{3}\otimes T^{(3,3)}_{13} \\
&+[C-\frac{1}{3}(c_{(\theta_{1}+\theta_{3})}+c_{\theta_{1}}+c_{\theta_{3}})+\frac{\mathrm{i}}{3}(s_{(\theta_{1}+\theta_{3})}-s_{\theta_{1}}-s_{\theta_{3}})] T^{(3,3)}_{12}\otimes I_{3} \\
&+[C-\frac{1}{3}(c_{(\theta_{2}+\theta_{4})}+c_{\theta_{2}}+c_{\theta_{4}})+\frac{\mathrm{i}}{3}(s_{(\theta_{2}+\theta_{4})}-s_{\theta_{2}}-s_{\theta_{4}})] T^{(3,3)}_{13}\otimes I_{3} \\
&+[C-\frac{1}{3}(c_{(\theta_{1}+\theta_{2})}+c_{(\theta_{1}+\theta_{3})}-c_{\theta_{1}}+c_{\theta_{2}}+c_{\theta_{3}})
+\frac{\mathrm{i}}{3}(s_{(\theta_{1}+\theta_{2})}+s_{(\theta_{1}+\theta_{3})}+s_{\theta_{1}}-s_{\theta_{2}}-s_{\theta_{3}})] T^{(3,3)}_{12}\otimes T^{(3,3)}_{12} \\
&+[C-\frac{1}{3}(c_{(\theta_{1}+\theta_{3})}+c_{(\theta_{3}+\theta_{4})}+c_{\theta_{1}}-c_{\theta_{3}}+c_{\theta_{4}})
+\frac{\mathrm{i}}{3}(s_{(\theta_{1}+\theta_{3})}+s_{(\theta_{3}+\theta_{4})}-s_{\theta_{1}}+s_{\theta_{3}}-s_{\theta_{4}})] T^{(3,3)}_{12}\otimes T^{(3,3)}_{13} \\
&+[C-\frac{1}{3}(c_{(\theta_{1}+\theta_{2})}+c_{(\theta_{2}+\theta_{4})}+c_{\theta_{1}}-c_{\theta_{2}}+c_{\theta_{4}})
+\frac{\mathrm{i}}{3}(s_{(\theta_{1}+\theta_{2})}+s_{(\theta_{2}+\theta_{4})}-s_{\theta_{1}}+s_{\theta_{2}}-s_{\theta_{4}})] T^{(3,3)}_{13}\otimes T^{(3,3)}_{12} \\
&+[C-\frac{1}{3}(c_{(\theta_{2}+\theta_{4})}+c_{(\theta_{3}+\theta_{4})}+c_{\theta_{2}}+c_{\theta_{3}}-c_{\theta_{4}})
+\frac{\mathrm{i}}{3}(s_{(\theta_{2}+\theta_{4})}+s_{(\theta_{3}+\theta_{4})}-s_{\theta_{2}}-s_{\theta_{3}}+s_{\theta_{4}})] T^{(3,3)}_{13}\otimes T^{(3,3)}_{13} ,
\end{split}
\end{equation}
where
\begin{equation}        \label{eq41}
\begin{split}
C=&\frac{1}{9}[c_{(\theta_{1}+\theta_{2}+\theta_{3}+\theta_{4})}
+\sum^{4}_{n=1}c_{\theta_{n}}+c_{(\theta_{1}+\theta_{2})}+c_{(\theta_{1}+\theta_{3})}+c_{(\theta_{2}+\theta_{4})}+c_{(\theta_{3}+\theta_{4})}
+\mathrm{i}(s_{(\theta_{1}+\theta_{2}+\theta_{3}+\theta_{4})}\\&
+\sum^{4}_{n=1}s_{\theta_{n}}-s_{(\theta_{1}+\theta_{2})}-s_{(\theta_{1}+\theta_{3})}-s_{(\theta_{2}+\theta_{4})}-s_{(\theta_{3}+\theta_{4})})].
\end{split}
\end{equation}
Hence, the Schmidt rank of $\mathcal{U}_{cd(3\otimes3)}$ is no more than nine.

\emph{Synthesis and Schmidt rank of $\mathcal{U}_{cd(M\otimes N)}$ on $\mathbb{C}^M\otimes \mathbb{C}^N$}.
When the $\mathcal{U}_{cd}$ is on $\mathbb{C}^M\otimes \mathbb{C}^N$, it can be decomposed as
\begin{equation}        \label{eq42}
\begin{split}
\mathcal{U}_{cd(M\otimes N)}=& (U_A\otimes U_B)\cdot \prod_{\stackrel{\tilde{a}=2,3,\cdots,M}{a=2,3,\cdots,N}}
                              e^{\mathrm{i}(\theta_i T_{1\tilde{a}}^{(M,3)}\otimes T_{1a}^{(N,3)})}\cdot(V_A\otimes V_B).
\end{split}
\end{equation}

Putting all the pieces together, and taking the trick ``absorbtion'' into account,  we find that $2M(N-1)$ GCX gates and $2M(N-1)+10$ single-partite $y$- and $z$-rotation-types are sufficient to implement a $\mathcal{U}_{cd(M \otimes N)}$ locally equivalent to a diagonal unitary.
It is noted that each $y$- or $z$-rotation-type has many free parameters in it when $M$, $N$ are larger.

We note that there are $MN-1$ linearly independent diagonal bases in $u(MN)$, and $\mathcal{U}_{cd(M\otimes N)}$ is equivalent to $\wedge(\triangle_{M\otimes N})=\prod_{\stackrel{\tilde{a}=2,3,\cdots,M}{a=2,3,\cdots,N}}e^{\mathrm{i}(\theta_i T_{1\tilde{a}}^{(M,3)}\otimes T_{1a}^{(N,3)})}$. Hence, $\wedge(\triangle_{M\otimes N})$ can be expanded as
\begin{equation}        \label{eq43}
\begin{split}
\wedge(\triangle_{M\otimes N})= &
\varepsilon I_{MN} + \sum_{k=1}^{N-1} \gamma_k  I_{M}\otimes T_{1k}^{(N,3)}+
\sum_{j=2}^{M-1} \delta_j T_{1j}^{(M,3)} \otimes I_{N}+
\sum_{j=2}^{M-1} \sum_{k=2}^{N-1}  \tau_{jk} T_{1j}^{(M,3)} \otimes T_{1k}^{(N,3)}.
\end{split}
\end{equation}
Based on Equation (\ref{eq43}), one can see that the Schmidt rank of a $\mathcal{U}_{cd(M\otimes N)}$ is no more than $MN$.

\section{Conclusion}\label{sec3}

Quantum circuit is the dominant paradigm for implementing and characterizing complex quantum computation, and the works are mainly focused on two-valued systems.
Utilizing Cartan decomposition technique, we have presented compact quantum circuits for implementing controlled-unitary gates with $A$ controlling on $\mathbb{C}^2\otimes \mathbb{C}^N$ and controlled-diagonal gates on $\mathbb{C}^M\otimes \mathbb{C}^N$, respectively.
We first showed the Cartan decomposition of Lie algebra $u(N)$ and reported that an arbitrary single-partite unitary can be implemented by three single-partite $y$- and $z$-rotation-types which are defined in this paper, and their appearance in decomposition of nonlocal gates is not previously studied, to our best knowledge.
Subsequently, we designed compact quantum circuits for implementing controlled-unitary gates with $A$ controlling on $\mathbb{C}^2\otimes \mathbb{C}^2$, $\mathbb{C}^2\otimes \mathbb{C}^3$, $\cdots$, $\mathbb{C}^2\otimes \mathbb{C}^N$ in terms of GCX gates and local single-partite rotation-types in detail.
The results indicate that $2(N-1)$ GCX gates together with  $6$ single-qubit rotations about the $y$- and $z$-axes, and $N+5$ single-partite $y$- and $z$-rotation-types are sufficient to implement a controlled-unitary gate with $A$ controlling on $\mathbb{C}^2\otimes \mathbb{C}^N$.
Lastly, we extended the program to gates on $\mathbb{C}^M\otimes \mathbb{C}^N$ which are locally equivalent to diagonal gates, and the quantum circuit is comprised of $2M(N-1)$ GCX gates and $2M(N-1)+10$ single-partite $y$- and $z$-rotation-types.

Based on Cartan decomposition, we have verified that expanding the diagonal form of the unitary is a way to get the upper bound of Schmidt rank of the controlled-unitary gate with $A$ controlling on $\mathbb{C}^2\otimes \mathbb{C}^N$ or the gate locally equivalent to a diagonal gate on $\mathbb{C}^M\otimes \mathbb{C}^N$, and such upper bound may be not tight.
The results showed that the controlled-unitary gate with $A$ controlling on $\mathbb{C}^2\otimes \mathbb{C}^N$ has Schmidt rank at most $2N$, whereas the one on $\mathbb{C}^2\otimes \mathbb{C}^2$ reaches the lower bound of 2, and the gate locally equivalent to a diagonal gate on $\mathbb{C}^M\otimes \mathbb{C}^N$ has Schmidt rank at most $MN$.
Although the bounds are not tight in general, the form of our decomposition involve special types of operators: the local operators in each term are either the identity or unitaries in low-dimensional subspaces. This explains that our decomposition in general has more terms than the Schmidt rank.

It is expected that the algorithm and technique employed here to implement controlled-unitary gates with $A$ controlling on $\mathbb{C}^2\otimes \mathbb{C}^N$  and the gates locally equivalent to diagonal gates on $\mathbb{C}^M\otimes \mathbb{C}^N$ could be useful for studying arbitrary multi-partite quantum computation in multi-valued systems.

\medskip

\section*{Acknowledgements} \par

This work is supported by the Fundamental Research Funds for the Central Universities under Grants FRF-TP-19-011A3, the National Natural Science Foundation of China under Grant No. 12004281 and 11704281.

\end{document}